\documentclass[aps,showpacs,twocolumn,superscriptaddress]{revtex4}
\usepackage{amssymb}
\usepackage{epsfig}
\usepackage{graphicx,amsmath}
\begin{document}

\title{Strongly correlated Fermi-systems: non-Fermi liquid behavior,
quasiparticle effective mass and their interplay}
\author{V.R. Shaginyan}\email{vrshag@thd.pnpi.spb.ru}
\affiliation{Petersburg Nuclear Physics Institute, RAS, Gatchina,
188300, Russia}\affiliation{Racah Institute of Physics, Hebrew
University, Jerusalem 91904, Israel}
\author{M.Ya. Amusia}\affiliation{Racah Institute
of Physics, Hebrew University, Jerusalem 91904, Israel}
\author{K.G. Popov}\affiliation{Komi Science Center, Ural Division,
RAS, Syktyvkar, 167982, Russia}

\begin{abstract}
Basing on the density functional theory of fermion condensation, we
analyze the non-Fermi liquid behavior of strongly correlated
Fermi-systems such as heavy-fermion metals. When deriving equations
for the effective mass of quasiparticles, we consider solids with a
lattice and homogeneous systems. We show that the low-temperature
thermodynamic and transport properties are formed by quasiparticles,
while the dependence of the effective mass on temperature, number
density, magnetic fields, etc gives rise to the non-Fermi liquid
behavior. Our theoretical study of the heat capacity, magnetization,
energy scales, the longitudinal magnetoresistance and magnetic
entropy are in good agreement with the remarkable recent facts
collected on the heavy-fermion metal $\rm YbRh_2Si_2$.
\end{abstract}
 \maketitle

\section{Introduction}

The Landau theory of Fermi liquids has a long history and remarkable
results in describing the properties of electron liquid in ordinary
metals and Fermi liquids of $^3$He type. The theory is based on the
Landau paradigm that elementary excitations determine the physics at
low temperatures. These excitations behave as quasiparticles, have a
certain effective mass $M^*$, which is independent of temperature
$T$, number density $x$, and magnetic field strength $B$ and is a
parameter of the theory \cite{land}. The discovery of strongly
correlated Fermi systems represented  by heavy-fermion (HF) metals
and 2D $^3$He exhibiting the non-Fermi liquid (NFL) behavior has
opened tremendous challenges in the modern condensed matter physics
\cite{senth,col,lohneysen,si,sach}. Facts collected on HF metals and
2D $^3$He demonstrate that the effective mass strongly depends on
$T$, $x$, $B$ etc, while $M^*$ itself can reach very high values or
even diverge \cite{lohneysen,si}. Such a behavior is so unusual that
the traditional Landau quasiparticles paradigm does not apply to it.

There is a common wisdom that quantum criticality, describing the
collective fluctuations of matter undergoing a second-order phase
transition at zero temperature, suppresses quasiparticles and thus
generates the NFL behavior, depending on the initial ground state,
either magnetic or superconductive
\cite{senth,col,lohneysen,si,sach}. Earlier, a concept of fermion
condensation quantum phase transition (FCQPT) preserving
quasiparticles and intimately related to the unlimited growth of
$M^*$, had been suggested  \cite{khs,ams}. Further studies show that
it is capable to deliver an adequate theoretical explanation of vast
majority of experimental results in different HF metals
\cite{volovik,obz,khodb}. In contrast to the Landau paradigm based
on the assumption that $M^*$ is a constant, in FCQPT approach $M^*$
strongly depends on $T$, $x$, $B$ etc. Therefore, in accord with
numerous experimental facts the extended quasiparticles paradigm is
to be introduced. The main point here is that the well-defined
quasiparticles determine as before the thermodynamic and transport
properties of strongly correlated Fermi-systems, while $M^*$ becomes
a function of $T$, $x$, $B$ etc \cite{obz,khodb,zph,ckz}. The FCQPT
approach had been already successfully applied to describe the
thermodynamic properties of such different strongly correlated
systems as $^3$He on one side and complicated heavy-fermion (HF)
compounds on the other side \cite{ckz,shag1,shag2,shag3,khodb,zph}.

In this letter, we analyze the non-Fermi liquid behavior of strongly
correlated Fermi systems using the density functional theory of
fermion condensation \cite{dft}. We derive equations for the
effective mass of quasiparticles in both homogeneous systems and
solids with a lattice, and show that extended quasiparticles
paradigm is valid, while the dependence of the effective mass on
$T$, $x$, $B$ etc gives rise to the NFL behavior. The obtained
results are illustrated with calculations of the thermodynamic and
transport functions of strongly correlated Fermi-systems. Possible
energy scales in these functions are discussed. We demonstrate that
our calculations of the heat capacity $C/T$, magnetization $M$,
energy scales, longitudinal magnetoresistance (LMR) and magnetic
entropy $S(B)$ are in good agreement with striking recent facts
collected on the HF metal $\rm YbRh_2Si_2$ \cite{oes,steg,geg}.

\section{Equation for the effective mass}

At first, consider HF liquid at $T=0$ characterized by the effective
mass $M^*$. Upon applying well-known Landau equation, we can relate
$M^*$ to the bare electron mass $M$ \cite{land, pfit}
\begin{equation}\label{MM*}
\frac{M^*}{M}=\frac{1}{1-N_0F^1(x)/3}.\end{equation} Here $N_0$ is
the density of states of a free electron gas, $M$ is the bare mass,
$x =p_F^3/3\pi^2$ is the number density, $p_F$ is the Fermi
momentum, and $F^1(x)$ is the $p$-wave component of Landau
interaction amplitude $F$. When at some critical point $x=x_c$,
$F^1(x)$ achieves certain threshold value, the denominator in Eq.
\eqref{MM*} tends to zero so that the effective mass diverges at
$T=0$ \cite{pfit}, and the system undergoes FCQPT. It follows from
Eq. \eqref{MM*} that beyond the critical point $x_c$, the effective
mass becomes negative. To avoid an unstable and physically
meaningless state with a negative effective mass, the system must
undergo a quantum phase transition at the quantum critical point
$x=x_c$, which is FCQPT \cite{ams,obz}. The asymmetrical phase
behind the quantum critical point is determined by
\begin{equation}\label{FCM}
\frac{\delta E}{\delta n({\bf p})}=\mu,\end{equation} here $E$ is
the ground state energy, $\mu$ is a chemical potential, and $n({\bf
p})$ is the occupation numbers of quasiparticles. The main result of
such reconstruction is that instead of Fermi step, we have $0\leq
n(p)\leq 1$ in certain range of momenta $p_i\leq p\leq p_f$.
Accordingly, the single-particle spectrum
\begin{equation}\label{SPS}\frac{\delta E}{\delta n({\bf
p})}=\varepsilon({\bf p}),\end{equation} in the above momenta
interval becomes flat, $\varepsilon({\bf p})=\mu$, and this state is
known as a fermi condensate (FC) state \cite{khs}. Due to the above
peculiarities of the $n({\bf p})$ function, FC state is
characterized by the superconducting order parameter $\kappa({\bf
 p})=\sqrt{n({\bf p})(1-n({\bf p}))}$.

To derive equation determining the effective mass, we employ the
density functional theory of superconducting state \cite{gross}. In
that case, the ground state energy $E$ becomes the functional of the
occupation numbers and the function of the number density $x$,
$E=E[n({\bf p}),x]$, while Eq. \eqref{SPS} gives the single-particle
spectrum \cite{dft}. Upon differentiating the both sides of Eq.
\eqref{SPS} with respect to ${\bf p}$ and after some algebra and
integration by parts, we obtain
\begin{equation}\label{EM}
\frac{\partial\varepsilon({\bf p})}{\partial {\bf p}}=\frac{{\bf
p}}{M}+\int F({\bf p},{\bf p}_1)\frac{\partial n({\bf
p}_1)}{\partial{\bf p}_1}\frac{d{\bf p}_1}{(2\pi)^3}.
\end{equation}
Here, $F({\bf p},{\bf p}_1)=\delta^2E/\delta n({\bf p})\delta n({\bf
p}_1)$ is the Landau amplitude. To calculate the derivative
$\partial\varepsilon({\bf p})/\partial {\bf p}$, we employ the
functional representation
\begin{eqnarray}\label{ENP}
E[n]&=&\int\frac{p^2}{2M}n({\bf p})\frac{d{\bf
p}}{(2\pi)^3}\nonumber
\\&+&\frac{1}{2}\int F({\bf p},{\bf p}_1)n({\bf p})n({\bf
p}_1)\frac{d{\bf p}d{\bf p}_1}{(2\pi)^6} +...
\end{eqnarray}
It is seen directly from Eq. \eqref{EM} that the effective mass is
given by the well-known Landau equation \begin{equation}\label{FLL}
\frac{1}{M^*} = \frac{1}{M}+\int \frac{{\bf p}_F{\bf p_1}}{p_F^3}
F({\bf p_F},{\bf p}_1)\frac{\partial n(p_1)}{\partial p_1}
\frac{d{\bf p}_1}{(2\pi)^3}.
\end{equation}
For simplicity, we ignore the spin dependencies. To calculate $M^*$
as a function of $T$, we construct the free energy $F=E-TS$, where
the entropy $S$ is given by
\begin{equation}
S=-2\int[n({\bf p})\ln (n({\bf p}))+(1-n({\bf p}))\ln (1-n({\bf
p}))]\frac{d{\bf p}}{(2\pi)^3},\label{FL3}
\end{equation}
which follows from general combinatorial reasoning \cite{land}.
Minimizing $F$ with respect to $n({\bf p})$, we arrive at the
Fermi-Dirac distribution,
\begin{equation} n({\bf p},T)=
\left\{1+\exp\left[\frac{(\varepsilon({\bf p},T)-\mu)}
{T}\right]\right\}^{-1}.\label{FL4} \end{equation} Due to the above
derivation, we conclude that Eqs. \eqref{EM} and \eqref{FLL} are
exact ones and allow us to calculate the behavior of both
$\partial\varepsilon({\bf p})/\partial {\bf p}$ and $M^*$ in the
vicinity of FCQPT where the well-defined quasiparticles determine
the low-temperature physics, while $M^*$ becomes a divergent
function of $T$, $B$ and $x$ \cite{obz,khodb,zph,ckz}. As we will
see this feature of $M^*$ forms the NFL behavior observed in
measurements on HF metals.

\section{Scaling behavior of the effective mass}

To avoid difficulties associated with the anisotropy generated by
the crystal lattice of solids, we study the universal behavior of
heavy-fermion metals using the model of the homogeneous
heavy-electron (fermion) liquid. The model is quite meaningful
because we consider the universal behavior exhibited by these
materials at low temperatures, a behavior related to power-law
divergences of quantities such as the effective mass, the heat
capacity, the magnetization, etc. These divergences and scaling
behavior of the effective mass, or the critical exponents that
characterize them, are determined by energy and momentum transfers
that are small compared to the Debye characteristic temperature and
momenta of the order of the reciprocal lattice cell length $a^{-1}$.
Therefore quasiparticles are influenced by the  crystal lattice
averaged over big distances compared to the length $a$.  Thus, we
can substitute the well-known jelly model for the lattice as it is
usually done, for example, in the fluctuation theory of second order
phase transitions.

\begin{figure}[!ht]
\begin{center}
\vspace*{-0.5cm}
\includegraphics [width=0.44\textwidth]{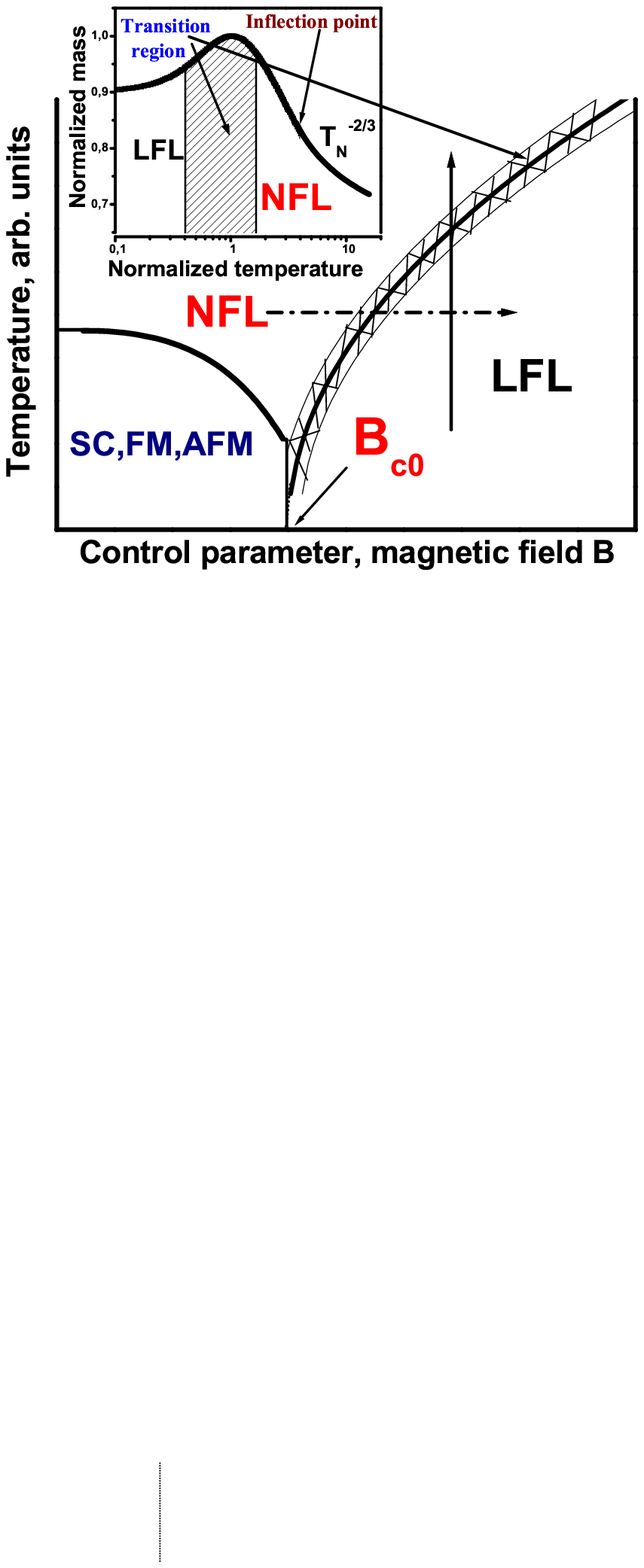}
\vspace*{-11.5cm}
\end{center}
\caption{ Schematic phase diagram of HF metals. $B_{c0}$ is magnetic
field at which the effective mass divergences. $\rm {SC,FM,AFM}$
denote the superconducting (SC), ferromagnetic (FM) and
antiferromagnetic (AFM) states, respectively. At $B<B_{c0}$ the
system can be in SC, FM or AFM states. The vertical arrow shows the
transition from the LFL regime to the NFL one at fixed $B$ along $T$
with $M^*$ depending on $T$. The dash-dot horizontal arrow
illustrates the system moving from NFL regime to LFL one along $B$
at fixed $T$. The inset shows a schematic plot of the normalized
effective mass versus the normalized temperature. Transition regime,
where $M^*_N$ reaches its maximum value $M^*_M$ at $T=T_M$, is shown
by the hatched area both in the main panel and in the inset. The
arrows mark the position of inflection point in $M^*_N$ and the
transition region.}\label{PHD}
\end{figure}

The schematic phase diagram of HF liquid is reported in Fig.
\ref{PHD}. Magnetic field $B$ is taken as the control parameter. In
fact, the control parameter can be pressure $P$ or doping (the
number density) $x$ etc as well. At $B=B_{c0}$, FC takes place
leading to a strongly degenerated state, where $B_{c0}$ is a
critical magnetic field, such that at $B>B_{c0}$ the system is
driven towards its Landau Fermi liquid (LFL) regime. In our simple
model $B_{c0}$ is a parameter. The FC state is captured by the
superconducting (SC), ferromagnetic (FM), antiferromagnetic (AFM)
etc. states lifting the degeneracy \cite{obz,khodb}. Below we
consider the HF metal $\rm YbRh_2Si_2$. In that case, $B_{c0}\simeq
0.06$ T ($B\bot c$) and at $T=0$ and $B<B_{c0}$ the AFM state takes
place \cite{geg}. At elevated temperatures and fixed magnetic field
the NFL regime occurs, while rising $B$ again drives the system from
NFL region to LFL one as shown by the dash-dot horizontal arrow in
Fig. \ref{PHD}. Below we consider the transition region when the
system moves from NFL regime to LFL one along the horizontal arrow
and it moves from LFL regime to NFL one along the vertical arrow as
shonw in Fig. \ref{PHD}. The inset to Fig. \ref{PHD} demonstrates
the behavior of the normalized effective mass $M^*_N=M^*/M^*_M$
versus normalized temperature $T_N=T/T_M$, where $M^*_M$ is the
maximum value that $M^*$ reaches at $T=T_M$. The $T^{-2/3}$ regime
is marked as NFL one since the effective mass depends strongly on
temperature. The temperature region $T\simeq T_M$ signifies the
crossover between the LFL regime with almost constant effective mass
and NFL behavior, given by $T^{-2/3}$ dependence. Thus temperatures
$T\sim T_M$ can be regarded as the crossover region between LFL and
NFL regimes.

To explore a scaling behavior of $M^*$, we write the quasiparticle
distribution function as $n_1({\bf p})=n({\bf p},T)-n({\bf p})$,
with $n({\bf p})$ is the step function, and Eq. \eqref{FLL} then
becomes
\begin{equation}
\frac{1}{M^*(T)}=\frac{1}{M^*}+\int\frac{{\bf p}_F{\bf
p_1}}{p_F^3}F({\bf p_F},{\bf p}_1)\frac{\partial n_1(p_1,T)}
{\partial p_1}\frac{d{\bf p}_1}{(2\pi) ^3}. \label{LF1}
\end{equation}
At FCQPT the effective mass $M^*$ diverges and Eq. \eqref{LF1}
becomes homogeneous determining $M^*$ as a function of temperature
\begin{equation}M^*(T)\propto T^{-2/3},\label{LTT}
\end{equation}
while the system exhibits the NFL behavior \cite{ckz,obz}. If the
system is located before FCQPT, $M^*$ is finite, at low temperatures
the system demonstrates the LFL behavior that is $M^*(T)\simeq
M^*+a_1T^2$, with $a_1$ is a constant, see the inset to Fig.
\ref{PHD}. Obviously, the LFL behavior takes place when the second
term on the right hand side of Eq. \eqref{LF1} is small in
comparison with the first one. Then, at rising temperatures the
system enters the transition regime: $M^*$ grows, reaching its
maximum $M^*_M$ at $T=T_M$, with subsequent diminishing. Near
temperatures $T\geq T_M$ the last "traces" of LFL regime disappear,
the second term starts to dominate, and again Eq. \eqref{LF1}
becomes homogeneous, and the NFL behavior restores, manifesting
itself in decreasing of $M^*$ as $T^{-2/3}$. When the system is near
FCQPT, it turns out that the solution of Eq. \eqref{LF1} $M^*(T)$
can be well approximated by a simple universal interpolating
function \cite{obz,ckz,shag2}. The interpolation occurs between the
LFL ($M^*\simeq M^*+a_1T^2$) and NFL ($M^*\propto T^{-2/3}$) regimes
thus describing the above crossover \cite{ckz,obz}. Introducing the
dimensionless variable $y=T_N=T/T_M$, we obtain the desired
expression \begin{equation}M^*_N(y)\approx
c_0\frac{1+c_1y^2}{1+c_2y^{8/3}}. \label{UN2}
\end{equation}
Here $M^*_N=M^*/M^*_M$ is the normalized effective mass,
$c_0=(1+c_2)/(1+c_1)$, $c_1$ and $c_2$ are fitting parameters,
parameterizing the Landau amplitude.

It is possible to transport Eq. \eqref{LF1} to the case of the
application of magnetic fields \cite{ckz,shag2}.  The application of
magnetic field restores the LFL behavior so that $M^*_M$ depends on
$B$ as
\begin{equation}\label{LFLB}
    M^*_M\propto (B-B_{c0})^{-2/3},
\end{equation} while
\begin{equation}\label{LFLT}T_M\propto \mu_B(B-B_{c0}),\end{equation}
where $\mu_B$ is the Bohr magneton \cite{ckz,shag2,obz}. Employing
Eqs. \eqref{LFLB} and \eqref{LFLT} to calculate $M^*_M$ and $T_M$,
we conclude that Eq. \eqref{UN2} is valid to describe the normalized
effective mass in external fixed magnetic fields with
$y=T/(B-B_{c0})$. On the other hand, Eq. \eqref{UN2} is valid when
the applied magnetic field becomes a variable, while temperature is
fixed $T=T_f$. In that case, as seen from Eqs. \eqref{LTT},
\eqref{UN2} and\eqref{LFLB}, it is convenient to rewrite both the
variable as $y=(B-B_{c0})/T_f$, and Eq. \eqref{LFLT} as
\begin{equation}\label{LFLf}\mu_B(B_M-B_{c0})\propto T_f.\end{equation}

It follows from Eq. \eqref{UN2} that in contrast to the Landau
paradigm of quasiparticles the effective mass strongly depends on
$T$ and $B$. As we will see it is this dependence that forms the NFL
behavior. It follows also from Eq. \eqref{UN2} that a scaling
behavior of $M^*$ near FCQPT point is determined by the absence of
appropriate external physical scales to measure the effective mass
and temperature. At fixed magnetic fields, the characteristic scales
of temperature and of the function $M^*(T,B)$ are defined by both
$T_M$ and $M^*_M$ respectively. At fixed temperatures, the
characteristic scales are $(B_M-B_{c0})$ and $M^*_M$. It follows
from Eqs. \eqref{LFLB} and \eqref{LFLT} that at fixed magnetic
fields, $T_M\to0$, and $M^*_M\to\infty$, and the width of the
transition region shrinks to zero as $B\to B_{c0}$ when these are
measured in the external scales. In the same way, it follows from
Eqs. \eqref{LTT} and \eqref{LFLf} that at fixed temperatures,
$(B_M-B_{c0})\to0$, and $M^*_M\to\infty$, and the width of the
transition region shrinks to zero as $T_f\to0$.

A few remarks are in order here. As we shall see in Subsections IV C
and D, in some cases temperature or magnetic field dependencies of
the effective mass or of other observable like the longitudinal
magnetoresistance do not have "peculiar points" like maximum. The
normalization are to be performed in the other points like the
inflection point shown in the inset to Fig. \ref{PHD}. Such a
normalization is possible since it is established on the internal
scales.

\section{Non-Fermi liquid behavior in $\rm YbRh_2Si_2$}

In what follows, we compute the effective mass and employ Eq.
\eqref{UN2} for estimations of considered values. To obtain the
effective mass $M^*(T,B)$, we solve Eqs. \eqref{FL4} and \eqref{LF1}
with special form of Landau interaction amplitude, see Refs.
\cite{ckz,obz} for details. Choice of the amplitude is dictated by
the fact that the system has to be in the FCQPT point, which means
that first two $p$-derivatives of the single-particle spectrum
$\varepsilon({\bf p})$ should equal zero. Since first derivative is
proportional to the reciprocal quasiparticle effective mass $1/M^*$,
its zero (where $1/M^*=0$ and the effective mass diverges) just
signifies FCQPT. Zeros of two subsequent derivatives mean that the
spectrum $\varepsilon({\bf p})$ has an inflection point at Fermi
momentum $p_F$ so that the lowest term of its Taylor expansion is
proportional to $(p-p_F)^3$ \cite{ckz}. After solution of Eq.
\eqref{LF1}, the obtained spectrum had been used to calculate the
entropy $S(B,T)$, which, in turn, had been recalculated to the
effective mass $M^*(T,B)$ by virtue of well-known LFL relation
$M^*(B,T)=S(B,T)/T$. We note that our calculations confirm the
validity of Eq. \eqref{UN2}.

\subsection{Heat capacity and the Sommerfeld coefficient}

Exciting measurements of $C/T\propto M^*$ on samples of the new
generation of $\rm YbRh_2Si_2$ in different magnetic fields $B$ up
to 1.5 T \cite{oes} allow us to identify the scaling behavior of the
effective mass $M^*$ and observe the different regimes of $M^*$
behavior such as the LFL regime, transition region from LFL to NFL
regimes, and the NFL regime itself. A maximum structure in
$C/T\propto M^*_M$ at $T_M$ appears under the application of
magnetic field $B$ and $T_M$ shifts to higher $T$ as $B$ is
increased. The value of $C/T=\gamma_0$ is saturated towards lower
temperatures decreasing at elevated magnetic field, where $\gamma_0$
is the Sommerfeld coefficient \cite{oes}.

It follows from Section III, that the transition region corresponds
to the temperatures where the vertical arrow in the main panel of
Fig. \ref{PHD} crosses the hatched area. The width of the region,
being proportional to $T_M\propto (B-B_{c0})$ shrinks,  $T_M$ moves
to zero temperature and $\gamma_0\propto M^*$ increases as $B\to
B_{c0}$. These observations are in accord with the facts \cite{oes}.

\begin{figure} [! ht]
\begin{center}
\vspace*{-0.5cm}
\includegraphics [width=0.49\textwidth]{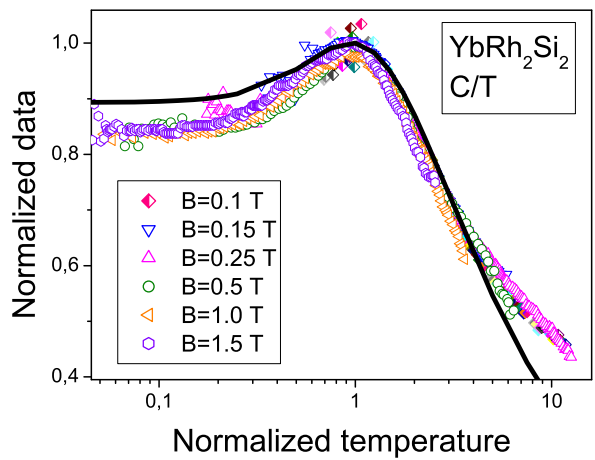}
\end{center}
\vspace*{-0.8cm} \caption{The normalized effective mass $M^*_N$
extracted from the measurements of the specific heat $C/T$ on $\rm
YbRh_2Si_2$ in magnetic fields $B$ shown in the left down corner
\cite{oes}. Our calculations are depicted by the solid curve tracing
the scaling behavior of $M^*_N$.}\label{fig2}
\end{figure}

To obtain the normalized effective mass $M^*_N$, the maximum
structure in $C/T$ was used to normalize $C/T$, and $T$ was
normalized by $T_M$. In Fig. \ref{fig2} the obtained $M^*_N$ as a
function of normalized temperature $T_N$ is shown by geometrical
figures, our calculations carried out as described above are shown
by the solid line. Figure \ref{fig2} reveals the scaling behavior of
the normalized experimental curves - the curves at different
magnetic fields $B$ merge into a single one in terms of the
normalized variable $y=T/T_M$. As seen from Fig. \ref{fig2}, the
normalized mass $M^*_N$ extracted from the measurements is not a
constant, as would be for a LFL, and shows the scaling behavior
given by Eq. \eqref{UN2} over three decades in normalized
temperature. The two regimes (the LFL regime and NFL one) separated
by the transition region, as depicted by the hatched area in the
inset to Fig. \ref{PHD}, are clearly seen in Fig. \ref{fig2}
illuminating good agreement between the theory and facts.

\subsection{ ``Average'' magnetization}

\begin{figure} [! ht]
\begin{center}
\vspace*{-0.5cm}
\includegraphics [width=0.49\textwidth]{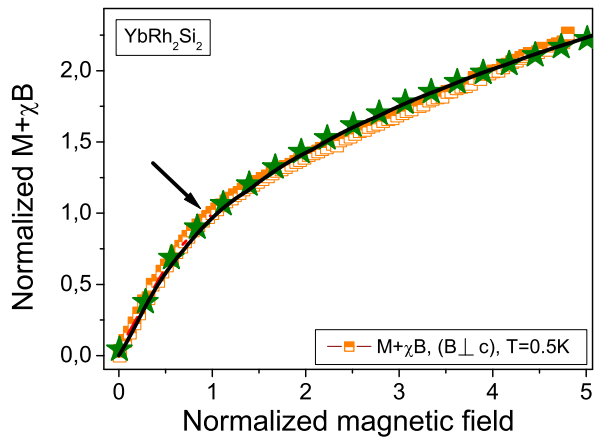}
\end{center}
\vspace*{-0.8cm} \caption{The field dependence of the normalized
``average'' magnetization $\widetilde{M}\equiv M+B\chi$ is shown by
squares and extracted from measurements collected on $\rm{
YbRu_2Si_2}$ \cite{steg}. The kink (shown by the arrow) is clearly
seen at the normalized field $B_N=B/B_k\simeq 1$. The solid curve
and stars (see text) represent our calculations.}\label{fig3}
\end{figure}

Consider now an ``average'' magnetization $\widetilde{M}\equiv
B\chi+M$ as a function of magnetic field $B$ at fixed temperature
$T=T_f$, where $\chi$ is the magnetic susceptibility and $M$ is the
magnetization,
\begin{equation}\label{CHIB}
M(B,T)=\int_0^B \chi(b,T)db,
\end{equation}
where the magnetic susceptibility is given by \cite{land}
\begin{equation}\label{CHI}
\chi(B,T)=\frac{\beta M^*(B,T)}{1+F_0^a}.
\end{equation}
Here, $\beta$ is a constant and $F_0^a$ is the Landau amplitude
related to the exchange interaction \cite{land}. In the case of
strongly correlated systems $F_0^a\geq -0.9$ \cite{pfit}. Therefore,
as seen from Eq. \eqref{CHI}, due to the normalization the
coefficients $\beta$ and $(1+F_0^a)$ drops out from the result, and
$\chi\propto M^*$. To obtain $\widetilde{M}$, we calculate $M^*$ as
a function of $B$ at fixed $T_f$. The obtained curves of
$\widetilde{M}$ exhibit energy scales separated by kinks at $B=B_k$.
As seen from Fig. \ref{fig3}, the kink is a crossover point from the
fast to slow growth of $\widetilde{M}$ at rising magnetic field. We
use $B_k$ and $\widetilde{M}(B_k)$ to normalize $B$ and
$\widetilde{M}$ respectively.

The normalized $\widetilde{M}$ vs the normalized field $B_N=B/B_K$
are shown in Fig. \ref{fig3}. Our calculations are depicted by the
solid line. The stars trace out our calculations of $\widetilde{M}$
with $M^*(y)$ extracted from the data $C/T$ shown in Fig.
\ref{fig2}. The calculation procedure deserves a remark here.
Namely, in that case $M^*$ depends on $y=T/T_M$ with $T_M$ is given
by Eq. \eqref{LFLT}. On the other hand, we can consider
$y=(B-B_{c0})/T_f$ as it is shown in Section III, and take the data
$C/T$ as a function of $y$.

It is seen from Fig. \ref{fig3} that our calculations are in good
agreement with the facts, and all the data exhibit the kink (shown
by arrow) at $B_N\simeq 1$ taking place as soon as the system enters
the transition region corresponding to the magnetic fields where the
horizontal dash-dot arrow in the main panel of Fig. \ref{PHD}
crosses the hatched area. Indeed, as seen from Fig. \ref{fig3}, at
lower magnetic fields $\widetilde{M}$ is a linear function of $B$
since $M^*$ is approximately independent of $B$. It follows from Eq.
\eqref{LFLB} that at elevated magnetic fields $M^*$ becomes a
diminishing function of $B$ and generates the kink in
$\widetilde{M}(B)$ separating the energy scales discovered in Ref.
\cite{steg}. Then, it seen from Eq. \eqref{LFLf} that the magnetic
field $B_k\simeq B_M$ at which the kink appears shifts to lower $B$
as $T_f$ is decreased.

\subsection{Longitudinal magnetoresistance}

\begin{figure} [! ht]
\begin{center}
\vspace*{-0.5cm}
\includegraphics [width=0.49\textwidth]{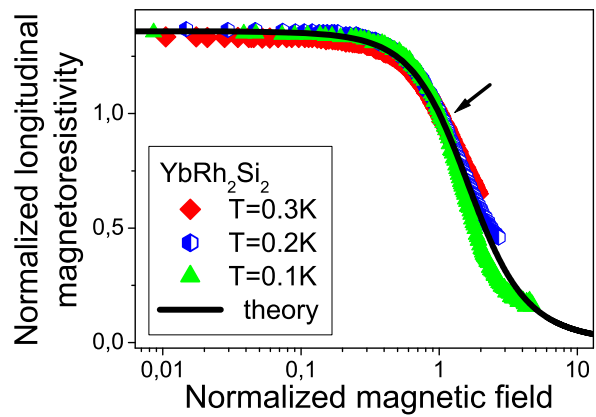}
\end{center}
\vspace*{-0.8cm} \caption{Magnetic field dependence of the
normalized (in the inflection point shown by the arrow, see text for
details) magnetoresistance $R^{\rho}_N$ versus normalized magnetic
field. $R^{\rho}_N$ was extracted from LMR of $\rm YbRh_2Si_2$ at
different temperatures \cite{steg} listed in the legend. The solid
line represents our calculations.}\label{fig4}
\end{figure}

Consider a longitudinal magnetoresistance (LMR)
$\rho(B,T)=\rho_0+AT^2$ as a function of $B$ at fixed $T_f$. In that
case, the classical contribution to LMR due to orbital motion of
carriers induced by the Lorentz force is small, while the
Kadowaki-Woods relation \cite{kadw,kwz}, $K=A/\gamma_0^2\propto
A/\chi^2=const$, allows us to employ $M^*$ to construct the
coefficient $A$ \cite{pla3}, since $\gamma_0\propto\chi\propto M^*$.
As a result, $\rho(B,T)-\rho_0\propto(M^*)^2$. Fig. \ref{fig4}
reports the normalized magnetoresistance
\begin{equation}\label{rn}
R_N^\rho(y)=\frac{\rho(y)-\rho_0}{\rho_{inf}}\propto (M_N^*(y))^2
\end{equation}
vs normalized magnetic field $y=B/B_{inf}$ at different
temperatures, shown in the legend. Here $\rho_{inf}$ and $B_{inf}$
are LMR and magnetic field respectively taken at the inflection
point marked by the arrow in Fig. \ref{fig4}.

The normalization procedure deserves a remark here. Namely, since
the magnetic field dependence of both the calculated $M^*$ and LMR
does not have "peculiar points" like maximums, the normalization
have been performed in the corresponding inflection points. To
determine the inflection point precisely, we first differentiate
$\rho(B,T)$ over $B$, find the extremum of derivative and normalize
the values of the function and the argument by their values in the
inflection point. Then, both theoretical (shown by the solid line)
and experimental (marked by the geometrical figures) curves have
been normalized by their inflection points, which also reveal the
universal behavior - the curves at different temperatures merge into
single one in terms of the scaled variable $y$ and show the scaling
behavior over three decades in the normalized magnetic field.

The transition region at which LMR starts to decrease is shown in
the inset to Fig. \ref{PHD} by the hatched area. Obviously, as seen
from Eq. \eqref{LFLf}, the width of the transition region being
proportional to $B_M\simeq B_{inf}$ decreases as the temperature
$T_f$ is lowered. In the same way, the inflection point of LMR,
generated by the inflection point of $M^*$ shown in the inset to
Fig. \ref{PHD} by the arrow, shifts to lower $B$ as $T_f$ is
decreased. All these observations are in excellent agreement with
the facts \cite{steg}.

\subsection{Magnetic entropy}

The evolution of the derivative of magnetic entropy $dS(B,T)/dB$ as
a function of magnetic field $B$ at fixed temperature $T_f$ is of
great importance since it allows us to study the scaling behavior of
the derivative of the effective mass $TdM^*(B,T)/dB\propto
dS(B,T)/dB$. While the scaling properties of the effective mass
$M^*(B,T)$ can be analyzed via MLR as we have shown in Subsection C.

\begin{figure} [! ht]
\begin{center}
\vspace*{-0.5cm}
\includegraphics [width=0.47\textwidth]{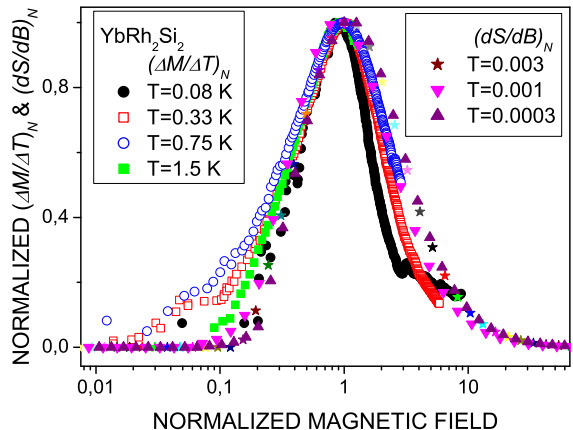}
\vspace*{-1.0cm}
\end{center}
\caption{Normalized magnetization difference divided by temperature
increment $(\Delta M/\Delta T)_N$ vs normalized magnetic field at
fixed temperatures (listed in the legend in the upper left corner)
is extracted from the facts collected  on $\rm YbRh_2Si_2$
\cite{geg}. Our calculations of the normalized derivative
$(dS/dB)_N\simeq (\Delta M/\Delta T)_N$ vs normalized magnetic field
are given at fixed dimensionless temperatures $T/\mu$ (listed in the
legend in the upper right corner). All the data are shown in the
geometrical figures depicted in the legends.} \label{fig5}
\end{figure}

As seen from the consideration in Subsection C and from Eqs.
\eqref{UN2} and \eqref{LFLf}, at $y\leq 1$ the derivative
$-dM_N(y)/dy\propto y$ with $y=(B-B_{c0})/(B_{inf}-B_{c0})\propto
(B-B_{c0})/T_f$. We recall that the effective mass as a function of
$B$ does not have the maximum, see Subsection C. At elevated $y$ the
derivative $-dM_N(y)/dy$ possesses a maximum at the inflection point
and then becomes a diminishing function of $y$. Upon using the
variable $y=(B-B_{c0})/T_f$, we conclude that at decreasing
temperatures, the leading edge of the function $-dS/dB\propto
-TdM^*/dB$ becomes steeper and its maximum at
$(B_{inf}-B_{c0})\propto T_f$ is higher. These observations are in
quantitative agreement with striking measurements of the
magnetization difference divided by temperature increment, $-\Delta
M/\Delta T$, as a function of magnetic field at fixed temperatures
$T_f$ collected on $\rm YbRh_2Si_2$ \cite{geg}. We note that
according to the well-know thermodynamic equality $dM/dT=dS/dB$, and
$\Delta M/\Delta T\simeq dS/dB$.

To carry out a quantitative analysis of the scaling behavior of
$-dM^*(B,T)/dB$, we calculate as described above the entropy
$S(B,T)$ as a function of $B$ at fixed dimensionless temperatures
$T_f/\mu$ shown in the upper right corner of Fig. \ref{fig5}. This
figure reports the normalized $(dS/dB)_N$ as a function of the
normalized magnetic field. The function $(dS/dB)_N$ is obtained by
normalizing $(-dS/dB)$ by its maximum taking place at $B_M$, and the
field $B$ is scaled by $B_M$. The measurements of $-\Delta M/\Delta
T$ are normalized in the same way and depicted in Fig. \ref{fig5} as
$(\Delta M/\Delta T)_N$ versus normalized field. It is seen from
Fig. \ref{fig5} that our calculations are in excellent agreement
with the facts and both the experimental functions $(\Delta M/\Delta
T)_N$ and the calculated $(dS/dB)_N$ ones show the scaling behavior
over three decades in the normalized magnetic field.

\section{Summary}

We have analyzed the non-Fermi liquid behavior of strongly
correlated Fermi systems using the density functional theory of
fermion condensation and derived equations for the effective mass of
quasiparticles in both homogeneous systems and solids with a
lattice, and showed that extended quasiparticles paradigm is
strongly valid, while the dependence of the effective mass on
temperature, number density, applied magnetic fields etc gives rise
to the NFL behavior. The obtained results are illustrated with
calculations of the thermodynamic and transport functions of
strongly correlated Fermi-systems. Possible energy scales in these
functions are discussed. We have demonstrated that our comprehensive
theoretical study of the heat capacity, magnetization, energy
scales, the longitudinal magnetoresistance and magnetic entropy are
in good agreement with the outstanding recent facts collected on the
HF metal $\rm YbRh_2Si_2$.

\section{Acknowledgements}

This work was supported in part by the grants: RFBR No. 09-02-00056
and the Hebrew University Intramural Funds. VRS is grateful to the
Lady Davis Foundation for supporting his visit to the Hebrew
University of Jerusalem.

\end{document}